\documentclass{article}
\usepackage{latexsym}
\usepackage{graphicx}

\begin{document}

\begin{center}
\textbf {\LARGE Is a Plasma Diamagnetic?}
\end{center}

\begin{center}
W. Engelhardt\footnote{ Electronic address: 
wolfgangw.engelhardt@t-online.de \\}  

\end{center}

\begin{center}
Max-Planck-Institut f\"{u}r Plasmaphysik, IPP-Euratom 
Association\footnote{
The content of this work is the sole responsibility of 
the author. In particular, the views expressed therein are not to be 
construed as being official and do not necessarily reflect those of the 
European Commission or the Max-Planck-Gesellschaft.}, Garching, Germany
\end{center}

\vspace{.6 cm} 

\noindent \textbf{Abstract}

\noindent Classical plasmas in thermodynamic equilibrium should be neither para- nor 
diamagnetic due to the action of the Lorentz force. Magnetic confinement, 
however, is based on the observed diamagnetism of laboratory plasmas. The 
apparent paradox is investigated on the basis of the resistive 
magneto-hydrodynamic equations. It is found that, at least in simple plasma 
configurations, these equations do not permit a solution, i. e. the paradox 
cannot be resolved. It seems that the Lorentz force is a test-particle 
approximation which is not suitable to describe the interaction of moving 
particles in agreement with the conservation of energy.

\vspace{.6 cm} 

\noindent \textbf{R\'{e}sum\'{e}}

\noindent Les plasmas classiques en \'{e}quilibre thermodynamique ne devraient \^{e}tre ni para- ni diamagn\'{e}tiques \`{a} cause de l'action de la force de 
Lorentz. Pourtant le confinement magn\'{e}tique des plasmas en laboratoire 
est fond\'{e} sur leur diamagn\'{e}tisme observ\'{e}. Le paradoxe est 
explor\'{e} sur la base des \'{e}quations de la th\'{e}orie 
magn\'{e}to-hydrodynamique r\'{e}sistive. On trouve que ces \'{e}quations ne 
permettent pas une solution unique, en tout cas pour des configurations 
simples; c'est \`{a} dire qu'il n'est pas possible de r\'{e}soudre le paradoxe. 
Il semble que la force de Lorentz repr\'{e}sente seulement une 
approximation pour des particules singuli\`{e}res. Si l'on veut observer la conservation d'\'{e}nergie, il faut admettre que cette force n'est pas 
appropri\'{e}e pour d\'{e}crire exactement l'interaction des particules en 
mouvement.

\vspace{.6 cm} 

\noindent \textbf{Keywords:}

\begin{enumerate}
\item magnetized plasmas
\item plasma diamagnetism and paramagnetism
\item Bohr-Van Leeuwen theorem
\item magnetohydrodynamics, ideal and resistive
\item magnetic plasma confinement, ideal and slowly diffusing equilibrium
\item theta pinch
\item magnetic interaction of moving particles
\item Lorentz force, Maxwell's equations, and conservation of energy
\end{enumerate}

\vspace{0.6 cm}

\noindent \textbf{I Introduction }

\noindent Feynman claims in his lectures [1] that both dia- and paramagnetism are 
exclusively quantum mechanical effects. His argument is of a very general 
nature: Since the classical Lorentz force is perpendicular to the velocity 
of a charged particle, the energy ${m\,v^2} \mathord{\left/ {\vphantom 
{{m\,v^2} 2}} \right. \kern-\nulldelimiterspace} 2$ of the particle does not 
depend on the magnetic field. Two boxes with the same number of particles 
and the same temperature must consequently contain the same energy even if 
one of the boxes is placed in a magnetic field. If the particles would alter 
the magnetic field, for example diamagnetically, the magnetic field energy 
would change and the total energy could not be the same. For this reason 
neither para- nor diamagnetism can arise as long as it is only the Lorentz 
force which acts upon the particles.

On the other hand, hot fusion plasmas are governed by the Lorentz force 
alone, since quantum-mechanical effects are negligible. Because of the 
rotational direction of the gyrating particles these plasmas behave clearly 
diamagnetically which is the reason why they can be confined by magnetic 
fields. In tokamaks, for example, one observes magnetic field changes, when 
the confined plasma is heated by external sources. The change of the 
toroidal magnetic flux is regularly monitored with a so called `diamagnetic 
loop' [2]. It allows to determine the energy content of the plasma [3] and 
yields information that can be utilized to control the position of the 
plasma in the vacuum vessel [4].

It has been pointed out that an ideally confined plasma is not in any 
contact with material walls and, therefore, not in complete thermodynamic 
equilibrium. This seems to explain why diamagnetism may occur classically in 
special circumstances. A. Schl\"{u}ter [5] quoting N. Bohr shows how the 
diamagnetism disappears in a plasma which is placed in a homogeneous field 
and surrounded by reflecting walls (Fig. 1). 

\vspace{0.6 cm}
\begin{center}
\includegraphics{dia1.epsi} \\
Figure 1 Gyration of particles in a box with reflecting walls (homogeneous 
field)
\end{center}

\vspace{0.6 cm}

\noindent Each gyrating particle 
constitutes a magnetic moment which by superposition would result in a 
magnetic field opposing the external field (diamagnetism). There is, 
however, an additional opposite magnetic moment created by the particles 
which are reflected at the walls such that the net effect is zero. This 
example confirms Feynman's conjecture, but it is not a generally valid 
demonstration. 

A counter-example may be produced by considering an inhomogeneous field: A 
straight wire carries a current and is surrounded by a toroidal vessel 
containing a plasma (Fig. 2).
 
\vspace{0.6 cm}

\begin{center}
\includegraphics{dia2.epsi} 

Figure 2 Gyration of particles in a box with reflecting walls (inhomogeneous field) 

\end{center}

\vspace{0.6 cm}

\noindent In this case -- in addition to the magnetic 
moment of the gyrating particles -- we have also a current density in the 
plasma volume parallel to the external current which is due to the particle 
drift in the inhomogeneous field. This current will alter the externally 
applied field so that the field energy is changed by the presence of the 
particles. As a consequence the energetic state depends on whether the 
vessel is being placed in the inhomogeneous field or not, in contrast to 
Feynman's general conclusion. 

In this paper we analyze the apparent paradox by applying the ideal and 
resistive magneto-hydrodynamic (mhd) equations to simple plasma 
configurations, but we find ourselves unable to remove the contradiction. It 
turns out that there is an intrinsic inconsistency between Lorentz force, 
Maxwell's equations, and energy conservation. This leads us to the 
conclusion that the Lorentz force is a test-particle approximation which 
ignores the back-reaction on the field-producing magnet. In most instances 
this is justified to a high degree of accuracy, as the test-particle 
interacts with typically $10^{23}$ field-producing particles. When it comes 
to a plasma, however, the field produced by the gyrating particles cannot be 
neglected any longer and the Lorentz force turns out to be insufficient to 
describe the interaction of the particles in motion.

\vspace{0.6 cm} 

\noindent \textbf{II The Resistive and Ideal Magneto-Hydrodynamic Model of a Plasma}

\noindent The mhd-equations are derived from the Boltzmann-equation applied to an 
ionized gas which is subject to the action of the Lorentz force. Derivations 
are found in many text-books. Quoting from [6] we have from the momentum 
balance of electrons and ions in a fully ionized hydrogen plasma:
\begin{equation}
\label{eq1}
\vec {j}\times \vec {B}=\nabla p+m_i {\kern 1pt}n\frac{d\vec {v}}{dt}
\end{equation}
\begin{equation}
\label{eq2}
\vec {E}+\vec {v}\times \vec {B}=\frac{1}{e{\kern 1pt}n}\left( {\vec 
{j}\times \vec {B}} \right)-\frac{\nabla p_e }{e{\kern 1pt}n}+\eta {\kern 
1pt}\vec {j}
\end{equation}
Here we have put $n=n_i =n_e $ and omitted terms of the order of the mass 
ratio ${m_e } \mathord{\left/ {\vphantom {{m_e } {m_i }}} \right. 
\kern-\nulldelimiterspace} {m_i }$. Furthermore, heat conduction and 
viscosity are neglected. For sufficiently slow processes Amp\`{e}re's law 
holds in the form:
\begin{equation}
\label{eq3}
rot\,\vec {B}=\mu _0 {\kern 1pt}\vec {j}
\end{equation}
Faraday's law of induction is:
\begin{equation}
\label{eq4}
rot\,\vec {E}=-\frac{\partial \vec {B}}{\partial t}
\end{equation}
We need the equation of continuity for the particle density:
\begin{equation}
\label{eq5}
div\,\left(n\vec {v}\right)=-\frac{\partial n}{\partial t}
\end{equation}
and the equation for the internal energy:
\begin{equation}
\label{eq6}
\frac{f}{2}{\kern 1pt}\frac{dp}{dt}+\frac{f+2}{2}\,p\,div\,\vec {v}=S+\eta 
{\kern 1pt}\vec {j}^2
\end{equation}
with $p=n\left( {T_i +T_e } \right)$ and $f=3$ for a plasma with three 
degrees of freedom. $S$ is a power density which is added to the plasma, 
e.g., by electromagnetic radiation. Equation (\ref{eq6}) does not contain the 
magnetic field which is a consequence of the Lorentz force acting presumably 
on the particles. Apart from the Joule heating term the equation is the same 
as the one for an ideal gas in the absence of a magnetic field.

If the electron temperature is sufficiently high, the terms in (\ref{eq2}) and (\ref{eq6}) 
containing the resistivity $\eta $, which accounts for the momentum exchange 
between electrons and ions, may be dropped. The resulting system of 
equations is the model of ideal mhd being valid on a time-scale short 
compared to the electron-ion collision time.

\vspace{0.6 cm} 

\noindent \textbf{III A Linear Theta-Pinch Heated by Radiation}

\noindent We apply the ideal mhd-equations with $\eta =0$ to a linear Theta-Pinch in 
equilibrium, the external field of which is produced by a superconducting 
coil (Fig. 3). 

\vspace{0.6 cm}

\begin{center}
\includegraphics{dia3.epsi} 

Figure 3 Field and pressure distribution in a Theta-Pinch
\end{center} 

\vspace{0.6 cm}

\noindent The straight field lines are parallel to the $z$- axis. From 
(\ref{eq1}) and (\ref{eq3}) follows for the internal and external field components:
\begin{equation}
\label{eq7}
p\left( r \right)+\frac{B_i^2 \left( r \right)}{2{\kern 1pt}\mu _0 
}=\frac{B_e^2 }{2{\kern 1pt}\mu _0 }
\end{equation}
We assume that at time $t=0$ the plasma is heated by switching on a 
radiation source so that the pressure is increased. Because of (\ref{eq7}) the 
magnetic field must change and the plasma radius $a$ defined by $p\left( a 
\right)=0$ may be displaced. As long as the radiation source is sufficiently 
weak, the kinetic energy of the plasma motion is negligible compared to the 
thermal energy and (\ref{eq7}) still holds during the expansion of the plasma. The 
total magnetic flux inside the coil remains unchanged because of (\ref{eq4}) as the 
electric field vanishes at the surface of the superconductor:
\begin{equation}
\label{eq8}
\int\limits_0^a {\frac{\partial B_i }{\partial t}\,r\,dr+} \int\limits_a^b 
{\frac{\partial B_e }{\partial t}\,r\,dr=0} 
\end{equation}
We insert (\ref{eq2}) into (\ref{eq4}):
\begin{equation}
\label{eq9}
\frac{\partial }{\partial r}\left( {r\,v\,B_i } \right)+\frac{\partial B_i 
}{\partial t}\,r=0
\end{equation}
where $v$ denotes the radial component of the plasma velocity, and integrate 
from 0 to $a$. Together with (\ref{eq8}) we obtain an equation for the change of the 
external field due to the expansion velocity of the plasma edge:
\begin{equation}
\label{eq10}
\frac{1}{B_e }{\kern 1pt}\frac{dB_e }{dt}=\frac{1}{b^2-a^2}{\kern 
1pt}\frac{da^2}{dt}
\end{equation}
An equation for the divergence of the velocity field follows from (\ref{eq6}):
\begin{equation}
\label{eq11}
v\,\frac{\partial p}{\partial r}+\frac{\partial p}{\partial t}=\frac{2{\kern 
1pt}S}{f}-\frac{f+2}{f}\frac{p}{r}\frac{\partial \left( {r\,v} 
\right)}{\partial r}
\end{equation}
by elimination of the time derivative of the pressure with (\ref{eq7}) and (\ref{eq9}):
\begin{equation}
\label{eq12}
\frac{1}{r}{\kern 1pt}\frac{\partial \left( {r\,v} \right)}{\partial 
r}=\frac{2{\kern 1pt}\mu _0 S-f{\kern 1pt}B_e {\left( {dB_e } \right)} 
\mathord{\left/ {\vphantom {{\left( {dB_e } \right)} {dt}}} \right. 
\kern-\nulldelimiterspace} {dt}}{f{\kern 1pt}B_e^2 -\left( {f-2} \right)\mu 
_0 {\kern 1pt}p}
\end{equation}
Integration from 0 to $a$ results in a second equation for the boundary 
velocity
\begin{equation}
\label{eq13}
{\kern 1pt}\frac{da^2}{dt}=\int\limits_0^a {\frac{4{\kern 1pt}\mu _0 
S-f{\kern 1pt} {\left( {dB_e^2 } \right)} \mathord{\left/ {\vphantom 
{{\left( {dB_e^2 } \right)} {dt}}} \right. \kern-\nulldelimiterspace} 
{dt}}{f{\kern 1pt}B_e^2 -\left( {f-2} \right)\mu _0 {\kern 1pt}p}} \,r\,dr
\end{equation}
which yields together with (\ref{eq10}) an equation for the change of the external 
field due to the applied heating power:
\begin{equation}
\label{eq14}
\frac{dB_e^2 }{dt}\left( {b^2-a^2+\int\limits_0^a {\frac{2{\kern 
1pt}f\,r\,dr}{f-\left( {f-2} \right){\mu _0 {\kern 1pt}p} \mathord{\left/ 
{\vphantom {{\mu _0 {\kern 1pt}p} {B_e^2 }}} \right. 
\kern-\nulldelimiterspace} {B_e^2 }}} } \right)=\int\limits_0^a 
{\frac{8{\kern 1pt}\mu _0 {\kern 1pt}S\,r\,dr}{f-\left( {f-2} \right){\mu _0 
{\kern 1pt}p} \mathord{\left/ {\vphantom {{\mu _0 {\kern 1pt}p} {B_e^2 }}} 
\right. \kern-\nulldelimiterspace} {B_e^2 }}} 
\end{equation}
The task is now to solve (\ref{eq11}) with the velocity as given by (\ref{eq12}) inside a 
moving boundary as described by (\ref{eq10}) and (\ref{eq14}). The boundary conditions are:

\[
\left[ {{\partial p} \mathord{\left/ {\vphantom {{\partial p} {\partial r}}} 
\right. \kern-\nulldelimiterspace} {\partial r}} \right]_{r=0} =0\;,\quad 
p\left( a \right)=0\,,\quad v\left( 0 \right)=0.
\]
As initial condition we may choose an arbitrary pressure profile $p\left( 
{r,{\kern 1pt}0} \right)$. If $p\left( a \right)=0$ is to hold at all times, 
we must require that the heating source $S$ vanishes at the plasma boundary. 
For the sake of simplicity we choose 
\begin{equation}
\label{eq15}
S\left( {r,t} \right)=\alpha \,p\left( {r,t} \right)
\end{equation}
where $\alpha $ is a constant.

We introduce dimensionless variables:
\begin{eqnarray}
r^2=x\,a^2\left( \tau \right)\;,\quad 0\le x\le 1\;,\quad 
t=\frac{f}{2{\kern 1pt}\alpha}\,\tau \quad \quad \quad \quad \quad \nonumber \\ \label{eq16}
\\
 p=\frac{B_e^2 }{2{\kern 1pt}\mu _0 }{\kern 1pt}\frac{\beta }{1+\delta 
\,\beta }\;,\quad \delta =\frac{f-2}{2{\kern 1pt}f}\;,\quad v=\frac{\alpha 
\,a^2}{r\,f}\left( {u\left( {x{\kern 1pt},\,\tau } 
\right)+\frac{x}{a^2}{\kern 1pt}\frac{da^2}{d\tau }} \right) \nonumber 
\end{eqnarray}
The transformation rules are:
\begin{equation}
\label{eq17}
\frac{1}{r}{\kern 1pt}\frac{\partial }{\partial r}=\frac{2}{a^2}{\kern 
1pt}\frac{\partial }{\partial x}\;,\quad \frac{\partial }{\partial 
t}=\frac{2{\kern 1pt}\alpha }{f}\left( {\frac{\partial }{\partial \tau 
}-\frac{x}{a^2}\frac{da^2}{d\tau }\,\frac{\partial }{\partial x}} \right)
\end{equation}
Equations (\ref{eq11}) and (\ref{eq12}) transform into:
\begin{equation}
\label{eq18}
u\frac{\partial \beta }{\partial x}+\frac{\partial \beta }{\partial \tau 
}=\beta \left( {1-\left( {1-\delta } \right)\beta } \right)\left( {1+\delta 
\,\beta } \right)\left( {1-\frac{2{\kern 1pt}\delta }{B_e }\frac{dB_e 
}{d\tau }} \right)
\end{equation}
\begin{equation}
\label{eq19}
\frac{\partial u}{\partial x}=\beta \left( {\frac{1}{2}-\frac{\delta }{B_e 
}\frac{dB_e }{d\tau }} \right)-\frac{1}{a^2B_e }\frac{d\left( {a^2B_e } 
\right)}{d\tau }
\end{equation}
The boundary conditions following from (\ref{eq18}) and (\ref{eq16}) are:
\begin{eqnarray}
\frac{d\beta \left( {0{\kern 1pt},\,\tau } \right)}{d\tau} \!\!\!&=&\!\!\! \beta \left( 
{0{\kern 1pt},\,\tau } \right)\left[ {1-\left( {1-\delta } \right)\beta 
\left( {0{\kern 1pt},\,\tau } \right)} \right]\,\left[ {1+\delta \,\beta 
\left( {0{\kern 1pt},\,\tau } \right)} \right]\,\left( {1-\frac{2{\kern 
1pt}\delta }{B_e }{\kern 1pt}\frac{dB_e }{d\tau }} \right) \nonumber \\ \label{eq20} \\
\beta \left( {1{\kern 1pt},\,\tau } \right) \!\!\!&=&\!\!\!0 \;,\quad\quad
u\left( {0{\kern 1pt},\,\tau } \right) = u\left( {1{\kern 1pt},\,\tau } 
\right)=0 \nonumber
\end{eqnarray}
We combine (\ref{eq18}) and (\ref{eq19}) into a single equation by taking $u$ as the 
independent variable instead of $x$:
\begin{equation}
\label{eq21}
u\,\frac{\partial \beta }{\partial u}{\kern 1pt}\frac{\partial u}{\partial 
x}+\frac{\partial \beta }{\partial \tau }=\beta \left( {1-\left( {1-\delta } 
\right)\beta } \right)\left( {1+\delta \,\beta } \right)\left( 
{1-\frac{2{\kern 1pt}\delta }{B_e }\frac{dB_e }{d\tau }} \right)
\end{equation}
This is possible, since (\ref{eq19}) is an ordinary differential equation which does 
not depend on $x$ explicitly. Upon substitution of (\ref{eq19}) equation (\ref{eq21}) is an 
inhomogeneous quasilinear equation of first order which has the 
characteristic system:
\begin{eqnarray}
\frac{1}{\beta }\frac{d\beta }{d\tau }=\beta \left( 
{1-\left( {1-\delta } \right)\beta } \right)\left( {1+\delta \,\beta } 
\right)\left( {1-\frac{2{\kern 1pt}\delta }{B_e }\frac{dB_e }{d\tau }} 
\right) \nonumber \\ \label{eq22} \\
\frac{1}{u}\frac{du}{d\tau }=\beta \,\left( 
{\frac{1}{2}-\frac{\delta }{B_e }\frac{dB_e }{d\tau }} 
\right)-\frac{1}{a^2B_e }{\kern 1pt}\frac{d\left( {a^2B_e } \right)}{d\tau } 
\quad\quad \nonumber 
\end{eqnarray}
The initial condition on $u$ results from (\ref{eq19}) and (\ref{eq20}):
\begin{equation}
\label{eq23}
u\left( {x{\kern 1pt},\,0} \right)=\left( {\frac{1}{2}-\frac{\delta }{B_e 
\left( 0 \right)}\left[ {\frac{dB_e }{d\tau }} \right]_{\tau =0} } 
\right)\left( {\int\limits_0^x {\beta \left( {x{\kern 1pt},\,0} 
\right)dx-x\int\limits_0^1 {\beta \left( {x{\kern 1pt},\,0} \right)dx} } } 
\right)
\end{equation}
By solving the system of ordinary differential equations (\ref{eq22}) we obtain a 
relationship between $u$, $\beta $, and $\tau $ when we eliminate the 
initial profiles $u\left( {x{\kern 1pt},\,0} \right)$ and $\beta \left( 
{x{\kern 1pt},\,0} \right)$ with (\ref{eq23}). It may be inserted into (\ref{eq19}) in order 
to express $u$ and $\beta $ as functions of $x$ and $\tau $ by further 
integration.

It turns out, however, that the solution of (\ref{eq22}) does not satisfy the 
boundary conditions (\ref{eq20}) in general. In order to demonstrate this we choose 
for simplicity the special case $\delta =0$ corresponding to $f=2$. It will 
become obvious that the difficulty remains in the more physical case $f=3$ . 
As an initial $\beta $- profile we take $\beta \left( {x{\kern 1pt},\,0} 
\right)=1-x^2$. The initial $u$- profile becomes with (\ref{eq23}): $u\left( 
{x{\kern 1pt},\,0} \right)=x\,{\left( {1-x^2} \right)} \mathord{\left/ 
{\vphantom {{\left( {1-x^2} \right)} 6}} \right. \kern-\nulldelimiterspace} 
6$ . By elimination of $x$ we have:
\begin{equation}
\label{eq24}
u\left( {x{\kern 1pt},\,0} \right)=\frac{1}{6}\,\beta \left( {x{\kern 
1pt},\,0} \right)\sqrt {1-\beta \left( {x{\kern 1pt},\,0} \right)} 
\end{equation}
Integration of (\ref{eq22}) yields:
\begin{equation}
\label{eq25}
\frac{\beta }{1-\beta }=\frac{\beta \left( {x{\kern 1pt},\,0} 
\right)\,e^\tau }{1-\beta \left( {x{\kern 1pt},\,0} \right)}
\end{equation}
\begin{equation}
\label{eq26}
u=u\left( {x{\kern 1pt},\,0} \right)\frac{g\left( 0 \right)}{6{\kern 
1pt}g}\sqrt {1-\beta \left( {x{\kern 1pt},\,0} \right)\left( {1-e^\tau } 
\right)} \;,\quad g=a^2B_e 
\end{equation}
Elimination of the initial profiles from (24 - 26) leads to the result:
\begin{equation}
\label{eq27}
u=\frac{g\left( 0 \right)}{6{\kern 1pt}g}\frac{\beta \sqrt {1-\beta } 
\,e^{-\tau }}{\left( {1-\beta \left( {1-e^{-\tau }} \right)} \right)^2}
\end{equation}
In order to determine the time evolution of $g$, we differentiate (\ref{eq27}) with 
respect to $x$ and evaluate it at $\beta \left( {1{\kern 1pt},\,\tau} \right)=0$ 
with (\ref{eq19}) substituted:
\begin{equation}
\label{eq28}
\left[ {\frac{\partial \beta }{\partial x}} \right]_{x=1} =-\frac{6\,e^\tau 
}{g\left( 0 \right)}\frac{dg}{d\tau }
\end{equation}
We may also differentiate (\ref{eq18}) with respect to $x$ and evaluate it at $x=1$ 
imposing (\ref{eq20}):
\begin{equation}
\label{eq29}
\frac{\partial ^2\beta }{\partial x\,\partial \tau }=\frac{\partial \beta 
}{\partial x}\left( {1+\frac{1}{g}\frac{dg}{d\tau }} \right)\;,\quad x=1
\end{equation}
Integration with respect to time yields:
\begin{equation}
\label{eq30}
\left[ {\frac{\partial \beta }{\partial x}} \right]_{x=1} =\left[ 
{\frac{\partial \beta \left( {x{\kern 1pt},\,0} \right)}{\partial x}} 
\right]_{x=1} \frac{g\,e^\tau }{g\left( 0 \right)}
\end{equation}
Elimination of the slope of $\beta $ at the boundary from (\ref{eq30}) and (\ref{eq28}) 
yields a differential equation for $g$:
\begin{equation}
\label{eq31}
\frac{1}{g}\frac{dg}{d\tau }=-\frac{1}{6}\,\left[ {\frac{\partial \beta 
\left( {x{\kern 1pt},\,0} \right)}{\partial x}} \right]_{x=1} 
\end{equation}
from which $g\left( \tau \right)$ may be determined.

If we insert, however, (\ref{eq31}) into (\ref{eq19}) and integrate from the axis to the 
boundary we find with (\ref{eq20}):
\begin{equation}
\label{eq32}
\int\limits_0^1 {\beta \,dx} =-\frac{1}{3}\,\left[ {\frac{\partial \beta 
\left( {x{\kern 1pt},\,0} \right)}{\partial x}} \right]_{x=1} =\frac{2}{3}
\end{equation}
Obviously, this integral equation is only satisfied at $\tau =0$. At later 
times $\beta $ evolves according to (\ref{eq18}) from the initial profile to $\beta 
\left( {x{\kern 1pt},\,\infty } \right)\to 1$, so that (\ref{eq32}) cannot hold at 
all times.

In view of this result we come to the conclusion that the set of equations 
(1 - 6) has no solution in general which would satisfy the boundary 
conditions. As the problem of heating a Theta-Pinch plasma in equilibrium is 
physically well posed, it must have a solution in reality. Evidently nature 
``uses equations'' which are different from those formulated in (1 - 6).

\vspace{0.6 cm} 

\noindent 
\textbf{IV A Slowly Diffusing Theta-Pinch Equilibrium}

\noindent Inclusion of finite resistivity removes the conservation of flux inside the 
plasma, but it does not remedy the situation. Starting from an equilibrium 
(\ref{eq7}) with a function $p\left( r \right)$ the plasma should slowly diffuse and 
ultimately fill the entire volume inside the coil. Spitzer [6] gives an 
expression for the diffusion velocity:
\begin{equation}
\label{eq33}
\vec {v}_{D\eta } =-\frac{\eta \,\nabla p}{B^2}
\end{equation}
which is easily derived from (\ref{eq1}) and (\ref{eq2}) under the assumption of the 
magnetic field staying constant in time. He rightly remarks that this 
condition is only satisfied in the test-particle approximation ${\mu _0 
{\kern 1pt}p} \mathord{\left/ {\vphantom {{\mu _0 {\kern 1pt}p} {B^2\to 0}}} 
\right. \kern-\nulldelimiterspace} {B^2\to 0}$, but, nevertheless, he claims 
one paragraph below that (\ref{eq33}) is of general validity restricted only by the 
exclusion of inertial terms.

Let us assume that there is no heating source $S$, but field energy is 
dissipated into internal energy by the diffusion process. Equation (\ref{eq8}) 
remains unchanged as long as the plasma diffusion occurs inside a 
superconducting coil. We derive the total power balance by taking the scalar 
product of (\ref{eq2}) with $\vec {j}$ and by eliminating the triple product with 
(\ref{eq1}):
\begin{equation}
\label{eq34}
\vec {E}\cdot \vec {j}=\eta \,\vec {j}^2+\vec {v}\cdot \nabla p
\end{equation}
By comparison with (\ref{eq6}) we find:
\begin{equation}
\label{eq35}
\frac{f}{2}\frac{\partial p}{\partial t}+\frac{f+2}{2}\,div\left( {p\,\vec 
{v}} \right)=\vec {E}\cdot \vec {j}
\end{equation}
Introducing the Poynting vector with (\ref{eq3}) and (\ref{eq4}) we have:
\begin{equation}
\label{eq36}
\frac{f}{2}\frac{\partial p}{\partial t}+\frac{f+2}{2}\,div\left( {p\,\vec 
{v}} \right)\,+\frac{1}{\mu _0 }\,div\left( {\vec {E}\times \vec {B}} 
\right)+\frac{1}{2{\kern 1pt}\mu _0 }\frac{\partial B^2}{\partial t}=0
\end{equation}
Integration over the plasma volume up to the coil radius using Gauss' 
theorem yields:
\begin{equation}
\label{eq37}
\int\limits_0^a {\left( {\frac{f}{2}\frac{\partial p}{\partial 
t}+\frac{1}{2{\kern 1pt}\mu _0 }\frac{\partial B_i^2 }{\partial t}} 
\right)\,r\,dr} +\int\limits_a^b {\left( {\frac{1}{2{\kern 1pt}\mu _0 
}\frac{\partial B_e^2 }{\partial t}} \right)\,r\,dr=0} 
\end{equation}
as the surface integrals arising from the divergence terms vanish. For 
simplicity we have omitted the term with the kinetic energy ${m_i {\kern 
1pt}n\,v^2} \mathord{\left/ {\vphantom {{m_i {\kern 1pt}n\,v^2} 2}} \right. 
\kern-\nulldelimiterspace} 2$ as the diffusion velocity is very small 
compared to the thermal speed. Furthermore, in accordance with (\ref{eq3}) we have 
neglected the electrostatic field energy arising through the electric field 
component $E_r ={\left( {{\partial p_i } \mathord{\left/ {\vphantom 
{{\partial p_i } {\partial r}}} \right. \kern-\nulldelimiterspace} {\partial 
r}} \right)} \mathord{\left/ {\vphantom {{\left( {{\partial p_i } 
\mathord{\left/ {\vphantom {{\partial p_i } {\partial r}}} \right. 
\kern-\nulldelimiterspace} {\partial r}} \right)} {e\,n}}} \right. 
\kern-\nulldelimiterspace} {e\,n}$, which provides the confinement of the 
ions. This contribution is negligibly small in laboratory plasmas compared 
to thermal and magnetic field energy. The heating term $\eta \,\vec {j}^2$ 
does not appear in (\ref{eq37}) explicitly. The increase in internal energy must 
come from a decrease of the field energy as the system is energetically 
closed by the condition $\vec {E}=0$ at the superconducting surface.

We insert (\ref{eq2}) into (\ref{eq4}):
\begin{equation}
\label{eq38}
\frac{\partial }{\partial r}\left( {r\,v\,B_i +r\,\eta \,j} 
\right)+\frac{\partial B_i }{\partial t}\,r=0
\end{equation}
This equation must be solved together with the local power balance (\ref{eq6}):
\begin{equation}
\label{eq39}
\frac{f}{2}\left( {\frac{\partial p}{\partial t}+v\,\frac{\partial 
p}{\partial r}} \right)=\eta {\kern 
1pt}j^2-\frac{f+2}{2}\,p\,\frac{1}{r}\frac{\partial \left( {r\,v} 
\right)}{\partial r}
\end{equation}
The solutions for the pressure and the internal magnetic field must satisfy 
the force balance (\ref{eq7}). We write the velocity as the sum of Spitzer's 
diffusion velocity (\ref{eq33}) and a term accounting for the effect of a finite 
pressure which is not included in (\ref{eq33}):
\begin{equation}
\label{eq40}
r\,v=-\frac{r\,\eta \,j}{B_i }+r\,v_p =u_S +u_p 
\end{equation}
With the abbreviation $x=r^2$ equation (\ref{eq38}) reads:
\begin{equation}
\label{eq41}
\frac{\partial }{\partial x}\left( {u_p {\kern 1pt}B_i } 
\right)+\frac{1}{2}\frac{\partial B_i }{\partial t}=0
\end{equation}
and (\ref{eq39}) becomes together with (\ref{eq1}):
\begin{equation}
\label{eq42}
\left( {u_S +\frac{f}{f+2}\,u_p } \right)\,\frac{\partial p}{\partial 
x}+\frac{f}{\left( {f+2} \right)2}\frac{\partial p}{\partial t}+p\left( 
{\frac{\partial u_S }{\partial x}+\frac{\partial u_p }{\partial x}} 
\right)=0
\end{equation}
Substituting the internal field with (\ref{eq7}) into (\ref{eq41}) yields a second 
differential equation for the pressure:
\begin{equation}
\label{eq43}
u_p \,\frac{\partial p}{\partial x}+\frac{1}{2}\frac{\partial p}{\partial 
t}+2p\,\frac{\partial u_p }{\partial x}=\frac{B_e^2 }{2{\kern 1pt}\mu _0 
}\frac{\partial u_p }{\partial x}+\frac{1}{4{\kern 1pt}\mu _0 }\frac{dB_e^2 
}{dt}
\end{equation}
It is quite obvious that (\ref{eq42}) and (\ref{eq43}) cannot lead to the same solution. The 
characteristic system of the inhomogeneous first order equation (\ref{eq42}) is:
\begin{equation}
\label{eq44}
\frac{dx}{u_S +{f\,u_p } \mathord{\left/ {\vphantom {{f\,u_p } {\left( {f+2} 
\right)}}} \right. \kern-\nulldelimiterspace} {\left( {f+2} 
\right)}}=\frac{f+2}{f}\,2\,dt=-\frac{dp}{p\left( {{\partial u_S } 
\mathord{\left/ {\vphantom {{\partial u_S } {\partial x+{\partial u_p } 
\mathord{\left/ {\vphantom {{\partial u_p } {\partial x}}} \right. 
\kern-\nulldelimiterspace} {\partial x}}}} \right. 
\kern-\nulldelimiterspace} {\partial x+{\partial u_p } \mathord{\left/ 
{\vphantom {{\partial u_p } {\partial x}}} \right. 
\kern-\nulldelimiterspace} {\partial x}}} \right)}
\end{equation}
and that of (\ref{eq43}):
\begin{equation}
\label{eq45}
\frac{dx}{u_p }=2\,dt=\frac{\mu _0 {\kern 1pt}dp}{\left( {B_e^2 -2{\kern 
1pt}\mu _0 p} \right)\,{\partial u_p } \mathord{\left/ {\vphantom {{\partial 
u_p } {\partial x+{\left( {dB_e^2 } \right)} \mathord{\left/ {\vphantom 
{{\left( {dB_e^2 } \right)} {\left( {4\,dt} \right)}}} \right. 
\kern-\nulldelimiterspace} {\left( {4\,dt} \right)}}}} \right. 
\kern-\nulldelimiterspace} {\partial x+{\left( {dB_e^2 } \right)} 
\mathord{\left/ {\vphantom {{\left( {dB_e^2 } \right)} {\left( {4\,dt} 
\right)}}} \right. \kern-\nulldelimiterspace} {\left( {4\,dt} \right)}}}
\end{equation}
Suppose a solution $p\left( {x{\kern 1pt},\,t} \right)$ exists and is known. 
One may then calculate the velocity $u_p \left( {x{\kern 1pt},\,t} \right)$ 
by eliminating the time derivative of the pressure from (\ref{eq42}) and (\ref{eq43}):
\begin{equation}
\label{eq46}
u_p =-\int\limits_0^x {\frac{\left( {f+2} \right){\partial \left( {\mu _0 
{\kern 1pt}p\,u_S } \right)} \mathord{\left/ {\vphantom {{\partial \left( 
{\mu _0 {\kern 1pt}p\,u_S } \right)} {\partial x+f{\left( {dB_e^2 } \right)} 
\mathord{\left/ {\vphantom {{\left( {dB_e^2 } \right)} {\left( {4\,dt} 
\right)}}} \right. \kern-\nulldelimiterspace} {\left( {4\,dt} \right)}}}} 
\right. \kern-\nulldelimiterspace} {\partial x+f\,{\left( {dB_e^2 } \right)} 
\mathord{\left/ {\vphantom {{\left( {dB_e^2 } \right)} {\left( {4\,dt} 
\right)}}} \right. \kern-\nulldelimiterspace} {\left( {4\,dt} 
\right)}}}{f\,B_e^2 -\left( {f-2} \right)\mu _0 {\kern 1pt}p}} \,dx
\end{equation}
Now the characteristic systems (\ref{eq44}) and (\ref{eq45}) may be integrated. The four 
resulting families of characteristics must lie entirely in the surface 
$p\left( {x{\kern 1pt},\,t} \right)$. This is, however, not possible, unless 
at least $u_S \left( {x{\kern 1pt},\,t} \right)=0$ which is excluded at 
finite resistivity.

The so called `slowly diffusing equilibrium', which is intuitively expected 
in the case of finite resistivity, is not obtainable from the resistive 
mhd-equations when the pressure dependent term {\-} omitted by Spitzer in 
(\ref{eq33}) {\-}~is included. Only in the test-particle case ${\mu _0 {\kern 1pt}p} 
\mathord{\left/ {\vphantom {{\mu _0 {\kern 1pt}p} {B^2\to 0}}} \right. 
\kern-\nulldelimiterspace} {B^2\to 0}$ an approximate diffusion velocity may 
be obtained from (\ref{eq33}) at constant magnetic field. In general, the momentum 
balance (1, 2) and the power balance (\ref{eq6}) are inconsistent in conjunction 
with Maxwell's equations (3, 4). Again, we must conclude that nature uses a 
different set of equations or, more properly speaking, at least one of the 
laws of nature as codified in (1 - 6) must be incomplete. It should be noted 
that the discrepancy encountered cannot be resolved by inclusion of heat 
conduction. Its dependence on temperature is different from that of the 
resistivity so that a cancellation of terms is generally not possible.

\vspace{0.6 cm} 

\noindent \textbf{V Discussion and Conclusion }

\noindent The mathematical model of resistive mhd (1 - 6) is, of course, not an exact 
description of reality. It leaves out not only a number of well known 
effects such as heat conduction, viscosity, thermoelectricity, gravity, but 
it neglects also the finite mass of electrons, the difference of electron 
and ion density, the effects of quantum mechanics etc. The idealizations 
involved are common practice in the mathematical modelling of physical 
reality, but this should lead only to minor deviations of the predictions 
from the observations on basic features. In hydrodynamics similar 
approximations are made, but the results as derivable from the model 
equations are in reasonable agreement with observations.

In magneto-hydrodynamics the situation is different according to our 
analysis: The mathematical model does not permit a prediction in principle, 
as the equations are internally inconsistent and do not yield unique 
solutions. This is not acceptable even for an idealized model and one must 
find out the reason. The derivation of the hierarchy of equations follows 
the same principles as in hydrodynamics so that an inconsistency is not to 
be expected at first sight. If it arises nevertheless, it must have to do 
with an inconsistency in the basic interaction law between individual 
particles as it is described by Lorentz force and Maxwell's equations.

The validity of Maxwell´s equations has been sufficiently 
confirmed and cannot be put into doubt in the present context. This is also 
true for the energy principle. The correctness of the Lorentz force, when 
applied to individual particles, seems also to be sufficiently verified. It 
is, however, practically impossible in these experiments to measure the 
back-reaction of the orbiting particles on the field producing magnet. We 
can, therefore, not exclude that a term in the elementary force law, which 
either cancels or is negligible in test-particle experiments, has escaped 
the attention. At least one observation raises doubts:

The angular momentum vector of a negatively charged particle gyrating in a 
homogeneous field is parallel to the field vector. If many particles in a circular conductor rotate in the same direction, they form a current which 
produces a magnetic moment such that the total field at the center of the 
loop is decreased. In this case the conductor is unstable as it has a 
tendency to turn around an axis which lies in its plane when a perturbation 
occurs. This indicates that the particles in the conductor are in a higher 
energetic state than they would be, if the current of negative particles 
would flow opposite to their natural direction of gyration. It is, of 
course, well known that a current loop in a magnetic field has in addition 
to its self-energy an extra potential energy which is the negative scalar 
product of its magnetic moment with the field.

The Lorentz force, however, does not predict a difference in the energetic 
state of an individual particle regardless whether it rotates clockwise or 
counter-clockwise in a magnetic field. Suppose a charged particle is 
attached at the periphery of a rotatable disk, similar to `Feynman's 
paradox' [1]. As the Lorentz force points towards or away from the axis of the 
disk, no extra work is necessary to reverse the direction of rotation so 
that the particle's energy -- in contrast to the particles in a current 
loop -- is independent of the sense of rotation. In the Appendix we show 
explicitly how the Lorentz force is at variance with the conservation of 
energy when a single particle interacts with a superconducting magnet.

The comparison of a gyrating particle with a current loop seems to point to 
an inconsistency which is probably at the root of the discrepancy which we 
have found when the Lorentz force is applied to a plasma. Because of the 
large amount of particles involved, we need to know the correct force law 
describing their interaction, not only a test-particle approximation in an 
external field. The back-reaction on the field producing magnet, which, at 
sufficiently high pressure, is the plasma itself, cannot be neglected any 
longer. If the Lorentz force would be complemented by a suitable term making 
the energetic state of a particle dependent on whether it is in a magnetic 
field or not, the equation of the internal energy (\ref{eq6}) would be altered and, hopefully, the discrepancy could be removed.

\vspace{0.6 cm} 

\noindent 
\noindent \textbf{Appendix }

\noindent A charged particle is attached at the periphery of a rotatable disk (Fig. 4) 
which is turned by a motor. 


\begin{center}
\includegraphics{dia4.epsi} 

Figure 4 Interaction of a charged particle with a superconducting magnet
\end{center} 


\noindent In a concentric superconducting ring flows a 
current which produces a magnetic field perpendicular to the plane of the 
disk. The particle's equation of motion is:

\setcounter{equation}{0} 
\renewcommand{\theequation}{A.\arabic{equation}} 

\begin{equation}
m\,\frac{d\vec {v}}{dt}=\vec {F}_m +q\left( {\vec {E}+\vec {v}\times \vec 
{B}} \right)
\end{equation}
The first term is the motor force acting on the particle via the mechanical 
fixture, the second term is the Lorentz force. The power balance is obtained 
by taking the scalar product of (A.1) with the velocity:
\begin{equation}
\frac{m}{2}\frac{d\vec {v}^2}{dt}=\vec {v}\cdot \vec {F}_m +q\,\vec {v}\cdot 
\vec {E}
\end{equation}
An electric field at the position of the particle is present when the 
current in the ring changes in time. We derive its tangential component from 
the vector potential of the superconductor:
\begin{equation}
E_{s\varphi } =-\frac{\partial A_{s\varphi } }{\partial t}=-\frac{\mu _0 
}{4{\kern 1pt}\pi }\frac{dI}{dt}\int\limits_0^{2\pi } {\frac{b\cos \varphi 
\,d\varphi }{\left( {b^2+a^2-2{\kern 1pt}a\,b\cos \varphi } 
\right) ^{\frac{1}{2}} }} 
\end{equation}
where $a$ and $b$ are the radii of the particle orbit and the superconductor, 
respectively.

The particle produces also a vector potential:
\begin{equation}
\vec {A}_p =\frac{\mu _0 {\kern 1pt}q}{4{\kern 1pt}\pi }\frac{\vec 
{v}}{\left| {\vec {x}-\vec {x}{\kern 1pt}'} \right|}
\end{equation}
and an electric field when it is accelerated:
\begin{equation}
\vec {E}_p =-\frac{\partial \vec {A}_p }{\partial t}=-\frac{\mu _0 {\kern 
1pt}q}{4{\kern 1pt}\pi }\frac{1}{\left| {\vec {x}-\vec {x}{\kern 1pt}'} 
\right|}\frac{d\vec {v}}{dt}
\end{equation}
The tangential component of this field is:
\begin{equation}
E_{p\varphi } =-\frac{\mu _0 {\kern 1pt}q}{4{\kern 1pt}\pi }\frac{1}{\left| 
{\vec {x}-\vec {x}{\kern 1pt}'} \right|}\left( {-\frac{dv_x }{dt}\sin 
\varphi +\frac{dv_y }{dt}\cos \varphi } \right)
\end{equation}
where $\vec {x}{\kern 1pt}'$ denotes the position of the particle. As it is 
compelled to move in tangential direction we have:
\begin{equation}
E_{p\varphi } =-\frac{\mu _0 {\kern 1pt}q}{4{\kern 1pt}\pi }\frac{1}{\left| 
{\vec {x}-\vec {x}{\kern 1pt}'} \right|}\,\left[ {\frac{dv_\varphi }{dt}\cos 
\left( {\varphi -\varphi {\kern 1pt}'} \right)+v_\varphi \frac{d\varphi 
{\kern 1pt}'}{dt}\sin \left( {\varphi -\varphi {\kern 1pt}'} \right)} 
\right]
\end{equation}
At the position of the superconducting ring the tangential field component 
of the particle is:
\begin{equation}
E_{p\varphi } =-\frac{\mu _0 {\kern 1pt}q}{4{\kern 1pt}\pi 
}\,\frac{\frac{dv_\varphi }{dt}\cos \left( {\varphi -\varphi {\kern 1pt}'} 
\right)+v_\varphi \frac{d\varphi {\kern 1pt}'}{dt}\sin \left( {\varphi 
-\varphi {\kern 1pt}'} \right)}{\left( {a^2+b^2-2{\kern 1pt}a\,b\cos \left( 
{\varphi -\varphi {\kern 1pt}'} \right)} \right)^{\frac{1}{2}}}\,
\end{equation}
Integration over the angle $\varphi $ from 0 to $2{\kern 1pt}\pi $ yields 
the loop voltage:
\begin{equation}
U_p =-\frac{\mu _0 {\kern 1pt}q}{4{\kern 1pt}\pi }\int\limits_0^{2\pi } 
{\frac{\frac{dv_\varphi }{dt}\cos \alpha +v_\varphi \frac{d\varphi {\kern 
1pt}'}{dt}\sin \alpha }{\left( {a^2+b^2-2{\kern 1pt}a\,b\cos \alpha } 
\right)^{\frac{1}{2}}}\,b\,d\alpha \;,\quad \alpha =\varphi -\varphi {\kern 
1pt}'} \,
\end{equation}
In the limit $a\ll b$ one obtains:
\begin{equation}
U_p \simeq -\frac{\mu _0 {\kern 1pt}q}{4}\frac{dv_\varphi 
}{dt}\frac{a}{b}\,
\end{equation}
Obviously, the particle works on the superconductor which carries a current:
\begin{equation}
\label{eq57}
U_p {\kern 1pt}I=-\frac{\mu _0 {\kern 1pt}q}{4}\frac{dv_\varphi 
}{dt}\frac{a}{b}\,I\,
\end{equation}
Its energy is increased or decreased depending on the direction of rotation:

\begin{equation}
W_{ps} =-\frac{\mu _0 {\kern 1pt}q\,a}{4{\kern 1pt}b}v_\varphi \,I\,
\end{equation}
The superconductor cannot sustain an electric field and must consequently 
compensate the applied voltage to zero by changing its current:
\begin{equation}
U_p =L\frac{dI}{dt}
\end{equation}
$L$ is its coefficient of self-induction which may be calculated by inserting 
into (A.13) the self-induced voltage $2{\kern 1pt}\pi \,b\,{\partial 
A_{s\varphi } } \mathord{\left/ {\vphantom {{\partial A_{s\varphi } } 
{\partial t}}} \right. \kern-\nulldelimiterspace} {\partial t}$:
\begin{equation}
2{\kern 1pt}\pi \,b\frac{\mu _0 }{4{\kern 1pt}\pi }\frac{dI}{dt}\left[ 
{\int\limits_0^{2\pi } {\frac{\cos \varphi \,b\,d\varphi }{\left( 
{r^2+b^2-2{\kern 1pt}r\,b\,\cos \varphi } \right)^{\frac{1}{2}}}} } \right]_{r\to b} 
=L\frac{dI}{dt}
\end{equation}
or:
\begin{equation}
L=\frac{\mu _0 {\kern 1pt}b^2}{2}\left[ {\int\limits_0^{2\pi } {\frac{\cos 
\varphi \,d\varphi }{\left( {r^2+b^2-2{\kern 1pt}r\,b\,\cos \varphi } 
\right)^{\frac{1}{2}}}} } \right]_{r\to b} 
\end{equation}
$L$ is proportional to the radius $b$. The factor of proportionality depends 
on the cross-section of the current ring and diverges logarithmically for a 
thin filament.

Inserting now (A.10) into (A.13) we find the connection between acceleration 
of the particle and current change in the superconductor:
\begin{equation}
\frac{dI}{dt}=-\frac{\mu _{0{\kern 1pt}} q\,a}{4\,b\,L}\frac{dv_\varphi 
}{dt}
\end{equation}
Substituting this into (A.3) we obtain the electric field at the position of 
the particle:
\begin{equation}
E_{s\varphi } =-\frac{\mu _0 }{4{\kern 
1pt}b}\frac{dI}{dt}\int\limits_0^{2\pi } {\frac{\cos \varphi \,b\,d\varphi 
}{\left( {a^2+b^2-2{\kern 1pt}a\,b\cos \varphi } 
\right)^{\frac{1}{2}}}\simeq } \frac{\mu _0 {\kern 
1pt}q\,a}{4\,b\,L}\frac{dv_\varphi }{dt}\frac{\mu _0 {\kern 
1pt}a}{4\,b}\;,\quad for\;a\ll b
\end{equation}
Inserting this into the power balance (A.2) of the motor yields finally:
\begin{equation}
\vec {v}\cdot \vec {F}_m =\frac{m}{2}\frac{dv_\varphi ^2 }{dt}-\left( 
{\frac{\mu _0 {\kern 1pt}q\,a}{4{\kern 1pt}b}} \right)^2\frac{1}{2{\kern 
1pt}L}\frac{dv_\varphi ^2 }{dt}
\end{equation}
The work done by the motor is apparently independent of the direction of 
rotation. To achieve, however, the same kinetic energy of the particle, more 
energy is necessary when there is no superconductor present. The additional 
term arising from the Lorentz force is independent of the magnetic field and 
vanishes proportional to $b^{-3}$ because of (A.15).

The interaction energy $W_{ps} $ (A.12) which depends on the direction of 
rotation and on the magnetic field is not accounted for in (A.18). It 
corresponds quantitatively to the potential energy $-\vec {\mu }\cdot \vec 
{B}$ of a magnetic moment in a magnetic field: The average current of the 
particle on its orbit multiplied with the enclosed area is $q\,a\,{v_\varphi 
} \mathord{\left/ {\vphantom {{v_\varphi } 2}} \right. 
\kern-\nulldelimiterspace} 2=\left| {\vec {\mu }} \right|$ and the magnetic 
field at the position of the particle as obtainable from $\vec {B}=rot\,\vec 
{A}$ is ${\mu _0 {\kern 1pt}I} \mathord{\left/ {\vphantom {{\mu _0 {\kern 
1pt}I} {2{\kern 1pt}b}}} \right. \kern-\nulldelimiterspace} {2{\kern 
1pt}b}$. One would expect that the motor must supply or gain this extra 
work, but the Lorentz force does not allow for it.

\vspace{0.6 cm} 

\noindent \textbf{Acknowledgments}

\noindent The author appreciates the time and effort spent by many colleagues to 
criticize this paper. He is particularly grateful for the continuing 
encouragement from Dr. K.-H. Steuer and impartial advice from Dr. O. Kardaun in an extensive and difficult discussion of the subject. The author is also indebted to Dr. D. S\"{u}nder who carefully read the manuscript and verified the derivations given in the paper. As a result the quality of the presentation could be improved considerably. 

\newpage

\noindent \textbf{References}

\begin{enumerate}
\item R. P. Feynman, R. B. Leighton, M. Sands, ``The Feynman Lectures on Physics'', Vol. II, 34 - 6 and 17 - 4, (Addison-Wesley Publishing Company, Reading, Massachusetts, 1964 ). 
\item John Wesson, Tokamaks, Section 10.2, (Clarendon Press, Oxford, 1987).
\item P. J. McCarthy, K. S. Riedel, O. J. W. F. Kardaun, H. D. Murmann, K. Lackner, and the ASDEX Team, Nuclear Fusion \textbf{31}, 1595 (1991).
\item O. Barana, A. Murari, F. Sartori and Contributors to the EFDA-JET Workprogramme, Nuclear Fusion \textbf{44}, 335 (2004).
\item A. Schl\"{u}ter, Annalen der Physik, \textbf{10}, 422 (1952).
\item L. Spitzer, Jr., Physics of Fully Ionized Gases, Second Edition, (Interscience Publishers, New York, 1962).
\end{enumerate}

\vspace{1.0cm}

\noindent \textbf{Figure captions}

Figure 1 Gyration of particles in a box with reflecting walls (homogeneous 
field)

Figure 2 Gyration of particles in a box with reflecting walls (inhomogeneous 
field)

Figure 3 Field and pressure distribution in a Theta-Pinch

Figure 4 Interaction of a charged particle with a superconducting magnet


\end{document}